\documentclass[]{emulateapj}
\usepackage{hyperref}
\usepackage{enumitem}
\hypersetup{
    colorlinks=true,
    linkcolor=blue,
    citecolor=blue,
    filecolor=blue,
    urlcolor=blue
}
\usepackage{natbib}

 \providecommand{\adsurl}[1]{\href{#1}{ADS}}
\usepackage{amsmath,amssymb}
\usepackage{mathptmx}
\usepackage{mathtools}
\usepackage{ctable}
\usepackage{multirow}
\usepackage{dcolumn}
\DeclareMathAlphabet{\pazocal}{OMS}{zplm}{m}{n}

\usepackage{graphicx}

\begin{document}

\title{Cold Clouds as Cosmic-Ray Detectors}
 \author{Shmuel Bialy$^{1 \ \star}$}
\affiliation{$^{1}$ Harvard Smithsonian Center for Astrophysics, 60 Garden st., Cambridge, MA, 02138 | sbialy@cfa.harvard.edu}
\slugcomment{Accepted for publication in Nature Communications Physics}

%

\begin{abstract}

Low energy cosmic-rays (CRs) are responsible for gas heating and ionization of interstellar clouds, which in turn introduces coupling to Galactic magnetic fields.
So far the CR ionization rate (CRIR) has been estimated using indirect methods, such as its effect on the abundances of various rare molecular species.
Here we show that the CRIR may be constrained 
from line emission of H$_2$ rovibrational transitions, excited by CRs.
We derive the required conditions for CRs to dominate line excitation, and show that CR-excited lines may be detected with the Very Large Telescope (VLT) over 8 hours integration.
Our method, if successfully applied to a variety of clouds at different Galactic locations
 will provide improved constraints on the spectrum of low energy CRs and their origins.

\end{abstract}


\section*{Introduction}
\label{sec: intro}

The ionization fraction of atomic and molecular clouds
is a primary factor in determining the gas evolution: it determines the efficiency of heating and cooling processes, drives the chemistry and molecule formation, and enables coupling to Galactic magnetic fields.
Ultraviolet (UV) radiation from starlight provides gas ionization, but this is restricted to localized regions, exposed to intense fluxes in the vicinity of massive stars.
For the bulk of the gas in the Galaxy, the
ionization is governed by cosmic-rays (CRs) (see \cite{Grenier2015} for a review)

It is the low energy CRs  ($E \ll$GeV) that are responsible for gas ionization in the interstellar medium (ISM), 
however, 
direct observations from Earth
may only probe high energy CRs.
Over the past few decades, the CR ionization rate (CRIR), denoted $\zeta$ (hereafter $\zeta$ is the total number of H$_2$ ionization per molecule, per sec, including both ionizations by CRs and by the secondary electrons produced by CR ionization), was estimated through 
observations of
various molecules and molecular ions in the ISM, such as OH, OH$^+$, H$_2$O$^+$, H$_3^+$, ArH$^+$, etc.
When combined with chemical models, these observations constrain the CRIR, yielding typical values ranging from
$\zeta = 10^{-17}$ to $10^{-15}$ s$^{-1}$ in dense and diffuse Galactic clouds \cite[][]{Guelin1982, VanderTak2000, Indriolo2012, Neufeld2017b, Bialy2019c, Gaches2019},
and up to $\zeta \approx 10^{-14}$ s$^{-1}$ in the Galactic center \cite[][]{LePetit2016} and in extragalactic sources \cite[][]{Indriolo2018}.

However, these determinations rely on the abundances of secondary species and depend on various model assumptions. For example, 
the gas density, 
the rate coefficients of the chemical reaction,
the fractional abundances of H$_2$, e$^-$, O, CO, etc.,
the number of clouds along the line-of-sight \cite[][]{Dalgarno2006, Indriolo2007b}.
Other indirect methods for inferring the CRIR include
the analysis of the thermal balance of dust and gas \cite[][]{Crapsi2007, Glassgold2012, Ivlev2019},
the effect on deuterium fractionation \cite[][]{Caselli1998, Williams1998, Kong2015, Shingledecker2016}, and through radio recombination lines \cite[][]{Sorochenko2010, Oonk2017} and synchrotron radiation \cite[][]{Yusef-Zadeh2013}.

The mass of molecular clouds in the ISM is strongly dominated by  
H$_2$.
The gas in molecular clouds is typically cold and the H$_2$ molecules reside mostly in their ground electronic, vibrational and rotational configuration.
However,	
H$_2$ rotational and vibrational transitions have been  previously
observed in shocked warm gas ($T \gtrsim 1000$ K), 
where the H$_2$ levels are thermally excited.
H$_2$ emission lines are also routinely
observed in bright photon-dominated regions (PDRs), in which the H$_2$ 
is excited by UV pumping.
These regions are exposed to abnormally high UV fluxes, $\chi \gg 1$, where $\chi$ is the radiation intensity normalized to the mean interstellar radiation field \cite[]{Draine1978}.
As we show below, for the more typical conditions of molecular gas in the ISM, i.e., cold ($T \lesssim100$ K) and quiescent ($\chi \approx 1$), 
CRs are expected to dominate H$_2$ excitation.

Numerical computations for H$_2$ excitation  by energetic electrons were presented by \cite{Tine1997} and \cite{Dalgarno1999a}.
Excitation by UV photons has been discussed by \cite{Black1987} and \cite{Sternberg1988, Sternberg1989a}, and the
excitation through the H$_2$ formation process, has been the focus of \cite{LeBourlot1995}, \cite{Tine2003}, and \cite{Islam2010}.

In this paper, 
we show that the CRIR may be determined 
through observations of line-emission from the main constituent of the cloud mass, H$_2$.
The H$_2$ rovibrational levels are excited by interactions with energetic electrons, which are produced by CR ionization.
As they radiatively decay they produce line emission in the infrared (IR) that is $\propto \zeta$.
We adopt an analytic approach 
to quantify the
conditions required for robust detection of H$_2$ lines that are excited by CRs:
(a) we consider the various line excitation mechanisms and their dependence on astrophysical parameters 
and derive the critical $\zeta/\chi$ above which CR excitation dominates line emission over UV and formation pumping, and (b)
consider the feasibility of line detection above the continuum with state of the art instruments.

\ctable[
caption = \mbox{H$_2$ line emission and excitation for excitation by cosmic-rays (CRs), ultraviolet (UV) radiation, and H$_2$ formation},
notespar,
label = table,
pos = t,
star
]{l l l l l l l l | l || l l l l}{
H$_2$ Rovibrational transitions. The upper and lower states $u$ and $l$ are charactarized have vibrational-rotational quantum numbers, $v_u=1$, $v_l=0$, and different $J_u$, $J_l$, as indicated.
For each transition, $\lambda_{ul}$,  $E_{ul}$ and $A_{ul}$ are the transition wavelength, energy, and Einstein coefficient for spontaneous decay.
$\alpha_{(u),l}\equiv A_{ul}/\sum_l A_{ul}$ is the 
 probability to decay to state $l$ given that state $u$ is excited.
$p_{u,({\rm cr})}$ is the probability for populating state $u$, per CR excitation \cite[]{Gredel1995}. These values are weakly sensitive to the temperature for $T \lesssim 50$ K.
$f_{ul,({\rm cr})} = p_{u,({\rm cr})}\alpha_{(u),l}E_{ul}/\bar{E}_{\rm (cr)}$ (with $\bar{E}_{\rm (cr)} \approx 0.486$ eV) 
is the normalized line brightness, relative to the total line brightness for pure CR excitation.
Correspondingly, $f_{ul,({\rm uv})}$ and $f_{ul,({\rm f})}$
are the normalized brightness for UV \cite[]{Sternberg1988} and H$_2$ formation \cite[]{LeBourlot1995} excitation (normalized to the total line brightness for pure UV/formation excitation, respectively).
For H$_2$ formation excitation, we consiedred three different H$_2$ formation models, $\phi=(1, 2, 3)$ \cite[]{Black1987}.
}{ 
\FL
Transition & $J_u$ & $J_l$ & $\lambda$ & $E_{ul}$ & $A_{ul}$ & 
$\alpha_{(u),l}$ & $p_{u,{\rm (cr)}}$ & $f_{ul,{\rm (cr)}} \quad \quad \quad$ & $f_{ul,{\rm (uv)}}$ & 
\multicolumn{3}{c}{$f_{ul,{\rm (f)}}(\%)$ }
\\
 &  &  & ($\mu$m) & (eV) & $(10^{-7} \ {\rm s^{-1}})$ & 
 &  & $(\%)$ & $(\%)$ & $\phi=1$ & $\phi=2$ & $\phi=3$
 \ML
(1-0)O(2) &0 & 2 &2.63 &0.47 &8.56 &1.00 &0.47 &45.29 & 0.94 & 0.10 & 0.27 & 0.66\NN 
(1-0)Q(2) &2 & 2 &2.41 &0.51 &3.04 &0.36 &0.47 &17.67 & 1.01 & 0.17 & 0.34 & 0.88\NN 
(1-0)S(0) &2 & 0 &2.22 &0.56 &2.53 &0.30 &0.47 &15.99 & 0.91 & 0.16 & 0.30 & 0.80\NN 
(1-0)O(4) &2 & 4 &3.00 &0.41 &2.91 &0.34 &0.47 &13.57 & 0.78 & 0.13 & 0.26 & 0.67\ML
(1-0)Q(1) &1 & 1 &2.41 &0.52 &4.30 &0.50 &0.014 &0.76 & 2.05 & 0.47 & 1.10 & 2.94\NN 
(1-0)S(1) &3 & 1 &2.12 &0.58 &3.48 &0.42 &0.014 &0.71 & 1.86 & 0.87 & 1.15 & 2.82\NN 
(1-0)O(3) &1 & 3 &2.80 &0.44 &4.24 &0.50 &0.014 &0.64 & 1.74 & 0.41 & 0.94 & 2.45\NN 
(1-0)Q(3) &3 & 3 &2.42 &0.51 &2.79 &0.33 &0.014 &0.50 & 1.31 & 0.61 & 0.84 & 1.96\NN 
(1-0)O(5) &3 & 5 &3.23 &0.38 &2.09 &0.25 &0.014 &0.28 & 0.74 & 0.35 & 0.46 & 1.09\LL 
}

\section*{Results}
\subsection*{Cosmic-ray Pumping}
\label{sec: signal}

We consider the emission of H$_2$ vibrational transitions from cold molecular clouds, where the vibrational levels are excited by penetrating CRs (and secondary electron).
As we discuss below, the line brightness is proportional to the CRIR, and thus may be used to constrain the CRIR inside clouds. Because radiative decay rates are high compared to the excitation rates, any excitation quickly decays
  back to the initial ground state before encountering the next excitation.
 Therefore, it is possible to separate the contribution from various excitation processes: 
 CR excitation, UV excitation, and excitation following H$_2$ formation, 
 (as discussed in the following subsections).
We focus on cold $T \lesssim 50$ K  gas typical of dense molecular cloud interiors. 
 In Methods we discuss warmer gas and the dependence of the line intensities on temperature.

 Assuming
that the H$_2$ reside in the ground 
vibrational $(v=0,J)$ states
and that each vibrational excitation is rapidly followed by radiative decay 
(see Methods),
the surface brightness of a transition line is
\begin{equation}
\label{eq: I_ul}
  I_{ul,{\rm (cr)}} = \frac{1}{4 \pi} g N_{\rm H_2} \zeta_{\rm ex}  \ p_{u,({\rm cr})} \alpha_{(u)l} E_{ul}  \ ,
\end{equation}
where
$u$ and $l$ denote the upper and lower energy states of the transition and
\begin{equation}
g(\tau) \equiv \frac{1-\mathrm{e}^{-\tau}}{\tau} \approx \frac{1-\mathrm{e}^{-0.9N_{22}}}{0.9N_{22}}
\end{equation}
 accounts for dust extinction in the infrared.
 $\tau = \sigma_{d} N$ is the the optical depth for dust extinction, 
 and $\sigma_d \approx 4.5 \times 10^{-23}$ cm$^2$
is the cross-section per hydrogen nucleus (the numeric value is an average over 2-3 $\mu$m \cite[][]{Draine2011}), where
$N_{\rm H_2}$ and $N\approx 2N_{\rm H_2}$ are the column densities of H$_2$ and hydrogen nuclei, and 
$N_{22} \equiv N_{\rm H_2}/(10^{22} \ {\rm cm^{-2}})$.
 In the limit $\tau \ll 1$, $g \rightarrow 1$ and $I_{ul,{\rm (cr)}} \propto N_{\rm H_2}$, i.e., the optically thin limit.
   In the limit $\tau \gg 1$,
  $I_{ul,{\rm (cr)}}$ saturates as
  $gN_{\rm H_2} \rightarrow 1/(2\sigma_d) = 1.1 \times 10^{22}$ cm$^{-2}$. This is the optically thick limit.
 For typical conditions, $N_{22}=1$, $\tau=0.9$, and $g \approx 0.66$. 
$\zeta_{\rm ex} \propto \zeta$ is the total excitation rate by CRs and by secondary electrons, and 
$p_{u,({\rm cr})}(T)$ is the probability per CR excitation to excite  level $u$, as determined by the interaction cross-sections for CRs and H$_2(v=0, J)$, assuming the rotational levels of H$_2(v=0, J)$ are thermalized \cite[]{Gredel1995}.
The factor
$\alpha_{(u),l} \equiv A_{ul}/\sum_l A_{ul}$
 is the probability to decay to state $l$ given
 state $u$ is excited,
 $A_{ul}$ is Einstein coefficient for radiative decay, and
$E_{ul}$ is the energy of the transition.
When cascade from high energy states is important, the level populations are coupled.
However, for CR excitation of the low rotational levels of $v=1$, direct impact dominates 
and the excitation rates simplify to $\zeta_{\rm ex} p_{u,({\rm cr})}$.
Values for $E_{ul}$, $\alpha_{(u),l}$, and $p_{u,({\rm cr})}$ are presented Table \ref{table}.
As discussed in Methods, in the case that the CRIR decreases with cloud depth, $\zeta_{\rm ex}$ and $\zeta$ represent the CR excitation and ionization rates in cloud interiors.

The total brightness in all the emitted lines is
\begin{align}
\label{eq: I_tot cr}
  I_{\rm tot,{\rm (cr)}} &=   \frac{1}{4 \pi}  gN_{\rm H_2} \varphi \zeta  \  \bar{E}_{\rm (cr)}  \\ 
 & = 3.6 \times 10^{-7} gN_{22} \zeta_{-16}  \ {\rm erg \ cm^{-2} \ s^{-1} \ str^{-1} }  \ \nonumber ,
\end{align} 
where $\zeta_{-16} \equiv \zeta/({\rm 10^{-16} \ s^{-1}})$,
 $\bar{E}_{\rm (cr)} \equiv \sum_{ul} E_{ul} p_{u,{\rm (cr)}} \alpha_{(u)l}\approx 0.486 \ {\rm eV}$ is the mean transition energy, and $\varphi \equiv \zeta_{ex}/\zeta \approx 5.8$
is the number of excitations per CR ionization (see Methods).
The brightness in each individual line may be written as
\begin{equation}
\label{eq: I_ul}
    I_{ul,{\rm (cr)}} = f_{ul,{\rm (cr)}} \ I_{\rm tot,{\rm (cr)}}  
\end{equation}
   where
\begin{equation}
\label{eq: f_ul}
    f_{ul,{\rm (cr)}} \equiv p_{u,{\rm (cr)}} \alpha_{(u)l} E_{ul}/\bar{E}_{\rm (cr)} \ ,
\end{equation}
is the relative emission brightness.

The $f_{ul,{\rm (cr)}}$ values for the brightest lines ($f_{ul,{\rm (cr)}}>0.1 \%$) are presented in Table \ref{table}.
The brightest lines (by far) are 
 the (1-0) O(2), Q(2), S(0), O(4) 
 transitions of para-H$_2$.
These transitions are strong because
the $(v,J)=(1,0), (1,2)$ states are efficiently populated by direct impact
 excitation from the ground (0,0) state, while other levels are populated by radiative cascade \cite[]{Gredel1995, LeBourlot1995}.
 Radiative cascade populates hundreds of levels, and thus the excitation efficiency for each individual level is low.
The ortho-H$_2$ lines (odd $J$)  are weak because the H$_2$ resides almost entirely in the para-H$_2$ ground state $(v,J)=(0,0)$, and  para-to-ortho conversion is inefficient.
The ortho-lines become important in warmer gas (see Methods).

\begin{figure*}[t]
	\centering
		\includegraphics[width=0.8\textwidth]{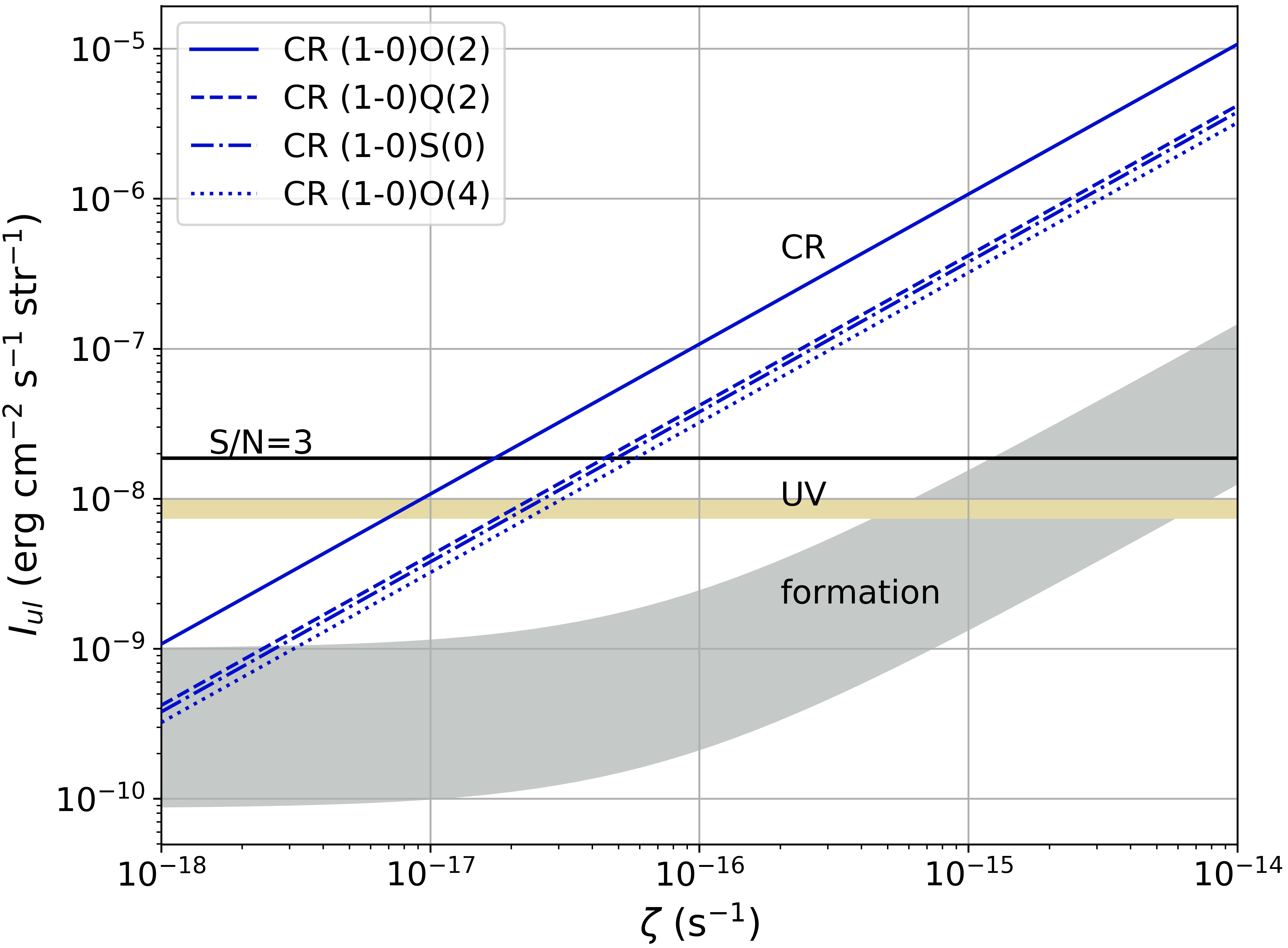} 
	\caption{
	Predicted energy surface brightness for the strongest  rovibrational transitions, emitted
	 from a molecular cloud of column density $N_{\rm H_{2}}=10^{22}$ molecules per cm$^2$ that is exposed to the mean interstellar ultraviolet (UV)  radiation field ($\chi=1$)\cite[]{Draine1978}, and to a cosmic-ray (CR) flux with an ionization rate $\zeta$.
	 Results are presented as functions of $\zeta$, and assuming 
	 either pure
	 CR excitation (four diagonal lines), pure UV excitation (yellow horizontal strip) and pure H$_2$ formation excitation (grey strip).
	For UV pumping, the strip includes the four transitions. For formation pumping, it also encompasses the three different formation models, $\phi=1, 2, 3$ \cite[]{Black1987}.
	The black horizontal line is the X-shooter sensitivity for a signal-to-noise ratio of 3, over  8 hrs integration time.
	For clouds with $\zeta >$ few 10$^{-17}$ s$^{-1}$, the line emission is dominated by CR excitation and may be detected with a night integration time.
		}
\label{fig: intensties}
\end{figure*}

In Fig.~\ref{fig: intensties} we show the strongest line brightness as a function of $\zeta$, for $N_{22}=\chi=1$.
The O(2) line is a factor of $\sim 3$ brighter than the other lines, Q(2), S(0), O(4), which are of comparable brightness.
This is because
there is equal probability for the excitation of 
both the  $(v,J)=(1,0)$ and $(v,J)=(1,2)$  levels (i.e., equal $p$),
but while the $(1,0)$ level is only allowed to decay to $(0,2)$, the $(1,2)$ level may decay to either $(0,2)$, $(0,0)$ or $(0,4)$.
This is reflected in Table \ref{table}, where
$\alpha=1$ for O(2) and $\alpha \approx 1/3$ for Q(2), S(0) and O(4).
For this reason,
  $\bar{E}_{\rm (cr)} \approx 0.486$ eV is so close to the energy of the O(2) line.

\subsection*{UV Pumping}
\label{sub: UV ex}

UV photons in the Lyman Werner  (LW) band (11.2-13.6 eV) excite the H$_2$ electronic states which cascade to the rovibrational levels of the ground electronic state.
This UV pumping is effective in the cloud envelopes.
With increasing cloud depth the radiation is attenuated by H$_2$ line absorption and dust absorption.
Assuming H$_2$ formation-destruction (by photodissociation) steady-state, where H$_2$ destruction leads to H formation and using the fact that the H$_2$ pumping and photodissociation rates are proportional,
 the surface brightness in all the lines may be written as
\begin{align}
\label{eq: I uv tot 1}
   I_{\rm tot, (uv)} &= 
   \frac{1}{4 \pi}  R n \frac{P_0}{D_0}  N_{\rm HI}  \bar{E}_{\rm (uv)} \ ,
\end{align}
see
\cite{Sternberg1988} for a derivation of a related quantity  (their Eq.~10).
In Eq.~(\ref{eq: I uv tot 1}), $R$ is the H$_2$ formation rate coefficient, $n$ is the gas density,
$D_0$ is the free-space H$_2$ photodissociation rate, $P_0 \approx 9 D_0$ is the UV pumping rate, 
and $\bar{E}_{\rm (uv)} \approx 1.82$ eV is the effective transition energy.
We derived $\bar{E}_{\rm (uv)}$ by comparing Eq.~(\ref{eq: I uv tot 1}) with \cite{Sternberg1988}'s computations of $N_{\rm HI}$ and $I_{\rm tot, (uv)}$.
The HI column density is
 \begin{equation}
\label{eq: N1_tot}
N_{\rm HI} = \frac{2 \langle \mu \rangle}{\sigma_g} \ln \left( \frac{\alpha G}{4 \langle \mu \rangle} + 1 \right) \ ,
\end{equation}
 where $\langle \mu \rangle \equiv \langle \cos(\theta) \rangle \approx 0.8$, $\sigma_g\equiv 1.9 \times 10^{-21} \tilde{\sigma}$ cm$^{2}$ is the dust absorption cross section over the LW band, per hydrogen nucleus, and $\tilde{\sigma}$ is the  cross-section in normalized units.
$\alpha \equiv D_0/(Rn)$ and $G \approx 3.0 \times 10^{-5} [9.9/(1+8.9 \tilde{\sigma})]^{0.37}$ is a self-shielding factor \cite[]{Sternberg2014, Bialy2016a}.
Eq.~(\ref{eq: N1_tot}) assumes slab geometry and irradiation by isotropic UV field of strength $\chi/2$ on each of side of the slab.
For beamed irradiation, multiply $\alpha G$ by 2.

For densities $n/\chi \gtrsim 20$ cm$^{-2}$, $\alpha G \lesssim 3.2$, and we may expand Eq.~(\ref{eq: N1_tot}), giving
$N_{\rm HI} = \alpha G/(2 \sigma_g)$, and 
\begin{align}
\label{eq: I uv tot 2}
   I_{\rm tot, (uv)} & \simeq  \frac{P_0 G}{8 \pi \sigma_g} \bar{E}_{\rm (uv)}   \\ \nonumber
   &\approx 9.6 \times 10^{-7}  \chi \ {\rm erg \ cm^{-2} \ s^{-1} \ str^{-1} } \ .
\end{align}
where in the second equality we used
$P_0=9D_0$, $D_0= 5.8 \times 10^{-11} \chi$ s$^{-1}$ and $\tilde{\sigma}=1$.
As long as $\chi/n<0.05$ cm$^3$, the brightness is independent of the density and the H$_2$ formation rate, and is proportional to the UV intensity, $\chi$.

Given $I_{\rm tot, (uv)}$, the brightness in an individual line excited by UV is 
\begin{equation}
\label{eq: I_ul uv}
    I_{ul,{\rm (uv)}} = f_{ul,{\rm (uv)}}I_{\rm tot, (uv)} \ , 
\end{equation}
where the relative emissions, $f_{ul,{\rm (uv)}}$, are determined by the Einstein radiative decay coefficients. 
The $f_{ul,{\rm (uv)}}$ values are given in Table \ref{table} based on \cite{Sternberg1988}.
They are of order 1\% and are much lower than the corresponding values for CR pumping. 
This is because UV pumping populates the levels through a cascade from electronic-excited-states, while for CR pumping, the levels are populated by direct impact excitation.
Fig.~\ref{fig: intensties} shows the resulting line brightness at $\chi=1$, and $N_{22}=1$.
Evidently, CR pumping dominates line emission for $\zeta_{-16}>0.1-0.2$.
More, generally, the ratio of emission arising from CR pumping relative to UV pumping is 
\begin{equation}
\label{eq: I_cr_I_uv}
    \frac{I_{ul, {\rm (cr)}}}{I_{ul, {\rm (uv)}}} = 0.38  \ \frac{\zeta_{-16}}{\chi}  \ \left( \frac{f_{ul,{\rm (cr)}}}{f_{ul,{\rm (uv)}}}  \right) \ gN_{22} \ ,
\end{equation}
and the critical $\zeta/\chi$ above which CR-pumping dominates is
\begin{equation}
\label{eq: zeta_crit}
        \left(  \frac{\zeta_{-16}}{\chi} \right)_{\rm c} = 2.7 \left( \frac{f_{ul,{\rm (cr)}}}{f_{ul,{\rm (uv)}}} g N_{22} \right)^{-1} \ .
\end{equation}
For $N_{22}=1$ ($g=0.66$), $(\zeta_{-16}/\chi)_{\rm c} \approx 0.08$ for O(2), and $\approx 0.2$ for Q(2), S(0), O(4).

\subsection*{Formation Pumping}
\label{sub form pump}
For each H$_2$ formed, a fraction of the binding energy is converted into level excitation.
It is useful to separate the line emission by H$_2$ formation pumping into a sum of two components, 
the contributions from the molecular core in which H$_2$ is destroyed by CRs, and from the outer envelopes where UV photons destroy H$_2$.
Assuming chemical steady state,
the H$_2$ formation is proportional to H$_2$ destruction and 
we get
\begin{align}
   \label{eq: I tot form 1}   
   I_{\rm tot, (f,core)} &= \frac{1}{4 \pi}  gN_{\rm H_2} y \zeta  \bar{E}_{\rm (f)}  \\ \nonumber
   &\approx 1.7 \times 10^{-7} \varphi_E gN_{22} y \zeta_{-16}  \ {\rm erg \ cm^{-2} \ s^{-1} \ str^{-1} } \ ,
   \\
   \label{eq: I tot form 2}
      I_{\rm tot, (f,env)} &= \frac{D_0 G}{8 \pi \sigma_g} \bar{E}_{\rm (f)} \\ \nonumber
   &\approx 7.6 \times 10^{-8}  \varphi_E \ \chi {\rm erg \ cm^{-2} \ s^{-1} \ str^{-1} } \ ,
\end{align}
for the inner core, and the outer envelopes, respectively.
Here we defined $\varphi_E \equiv \bar{E}_{\rm (f)}/(1.3 {\rm eV})$, where
$\bar{E}_{\rm (f)} \approx 1.3$ eV corresponds excitation of the $v=4$ level, as suggested by experiments \cite[]{Islam2010},
and the factor $y \approx 2$  accounts for additional removal of H$_2$ by H$_2^+$ in predominantly molecular gas \cite[][]{Bialy2015a}.
Eqs.~(\ref{eq: I tot form 1}, \ref{eq: I tot form 2}) have similar forms as Eqs.~(\ref{eq: I_tot cr}, \ref{eq: I uv tot 2}),
 as in the molecular core the H$_2$ removal rate is $\propto \zeta$, while in the outer envelopes removal is proportional to the UV pumping rate, $D_0 \propto P_0 \propto \chi$.
 The transition from core-to-envelope dominated formation pumping occurs when $\zeta/\chi$ is smaller than
\begin{equation}
\left(\frac{\zeta_{-16}}{\chi} \right)_{\rm crit} = 0.46 (gN_{22}y)^{-1} \ ,
\end{equation}
where $gN_{22}y$ is typically of order unity.

The surface brightness of each line is 
\begin{equation}
I_{ul, {\rm (f)}}=f_{ul,{\rm (f)}} I_{{\rm tot}, ({\rm f})} \ ,
\end{equation}
where $I_{\rm tot,(f)}=I_{\rm tot, (f,core)}+I_{\rm tot, (f,env)}$.
The $f_{ul,{\rm (f)}}$ values are determined by the formation excitation pattern, which is uncertain.
To illustrate the possible outcomes, we consider the three qualitatively different formation models, $\phi=1,2,3$ explored by \cite{Black1987} (see their Eqs.~2-4).
In Fig.~\ref{fig: intensties} we show the resulting line brightness for H$_2$ formation pumping, for the four lines and the three formation models  (grey strip), for $\chi=1$.
As  expected, when $\zeta_{-16} \gtrsim 1$, $I_{ul, {\rm (f)}} \propto \zeta$
as the cloud core dominates formation pumping, while when $\zeta_{-16} \lesssim 1$, $I_{ul, {\rm (f)}}$ is independent of $\zeta$.

However, importantly, in both limits H$_2$ formation pumping is never the dominant excitation mechanism.
When $\zeta/\chi \gg (\zeta/\chi)_{\rm crit}$,
\begin{equation}
\label{eq: I_cr_I_uv}
    \frac{I_{ul, {\rm (cr)}}}{I_{ul, {\rm (f)}}} \approx 2.2 (y\varphi_E)^{-1} \left( \frac{f_{ul,{\rm (cr)}}}{f_{ul,{\rm (f)}}}  \right) \ .
\end{equation}
Since $f_{ul,{\rm (cr)}} = 15-45 \%$ and
$f_{ul,{\rm (f)}} =0.1-1 \%$, CR pumping dominates line emission.
When $\zeta/\chi \ll (\zeta/\chi)_{\rm crit}$, although formation-pumping may be more important than CR-pumping,  
it remains sub-dominant compared to UV pumping, as can be seen by comparing Eqs.~(\ref{eq: I uv tot 2}) and (\ref{eq: I tot form 2}).
For formation-pumping to dominate over UV, the ratio $f_{ul, {\rm (f)}}/f_{ul, {\rm (uv)}}$ must be larger than $P_0/D_0 \approx 9$, which generally does not occur.


\subsection*{Continuum}
\label{subsub continuum background}
The astronomical source for continuum radiation in the wavelength of interest is dominated by light  reflected  from interstellar dust grains \cite[][]{Foster2006}.
Following \cite{Padoan2006}, in the optically thin limit, the specific intensity in the  K band is $I_{{\rm cont},\nu} \approx 8.0 \times 10^{-19} N_{22} \ {\rm erg \ cm^{-2}\  s^{-1}\ Hz^{-1} str^{-1} }$.
Integrating over a spectral bin $\Delta \nu$ (as the lines are narrow compared to $\Delta \nu$),
and 
multiplying by the optical depth correction function,  $g$, we get
\begin{equation}
\label{eq: I_cont}
    I_{\rm cont} =
     1.1 \times 10^{-8} gN_{22} R_4^{-1} \ {\rm erg \ cm^{-2}\  s^{-1} \ str^{-1} } \ ,
\end{equation}
where $R \equiv \nu/\Delta \nu$ is the resolving power, $R_4 \equiv R/10^4$, 
and where we used $\nu = 1.35 \times 10^{14}$ Hz corresponding to 2.2 $\mu$m.

Emission from small dust grains and polycyclic aromatic hydrocarbons (PAHs) heated by the interstellar UV field also contributes to the background continuum,  
\cite[][]{Draine2011} have calculated the emission spectrum assuming a realistic dust population composed of amorphous silicates and carbonaceous grains of various sizes \cite[]{Draine2007a} and including the effect of temperature fluctuations of small grains and PAHs.
At $\lambda=2-3$ $\mu$m, they find $\lambda I_{\lambda} \approx 2 \times 10^{-27} N I_{\rm UV}$ erg s$^{-1}$ str$^{-1}$ per H nucleus.
 Assuming $I_{\rm UV}=1$, $N = 10^{21}$ cm$^{-2}$ (at higher columns the UV flux is exponentially absorbed by dust), and integrating over a spectral bin we get 
 \begin{equation}
    I_{\rm cont, em} \approx 2 \times 10^{-10} R_4^{-1} \ {\rm erg \ cm^{-2} \ s^{-1} \ str^{-1}} \ ,
\end{equation}
Thus, at the wavelength of interest,
dust emission is subdominant compared to 
 scattered light.
 
\subsection*{Detectability}
\label{subsub: sensitivity}
For ground based observations, Earth sky thermal (and line) emission is typically the dominant noise source.
As a proof of concept we examine the detection feasibility with X-shooter on the Very Large Telescope (VLT) and focus on the S(0) and Q(2) lines (O(2) is blocked by the atmosphere and O(4) is outside X-shooter's range).
We assume that the lines are narrow and the source is extended.
For $\zeta_{-16}=N_{22}=1$,
the brightness of S(0) and Q(2) are $I=(3.8, 4.2) \times 10^{-8}$ erg cm$^{-2}$  s$^{-1}$ str$^{-1}$, respectively (Eq.~\ref{eq: I_tot cr},\ref{eq: I_ul}).
The estimated
signal-to-noise ratio (SNR) per pixel for 1 hour integration
with the 0.4'' slit ($R=11,600$),
 is 
$S/N=(0.29, 0.14)$ for S(0) and Q(2) respectively (see Methods).
For 8 hour integration, and
integrating along the slit (55 pixels), $S/N=(6.1, 2.9)$.

More generally, 
$S/N \propto \sqrt{t R \Delta \Omega}$,
where $\Delta \Omega$ is the instrument's field of view (FoV).
Nearby clouds extend over angles large compared to typical slit FoVs.
For example, the dark cloud Bernard 68 has an angular diameter $\approx 100''$ and 
$\Omega_{\rm B68} \approx 8,400$ arcsec$^2$, whereas the X-shooter slit FoV is only 11'' long and has $\Omega = 4.4$ arcsec$^2$.
Longer slits will achieve better SNR, but
the improvement is limited to a factor $\sqrt{9}$.
The achieved SNR  will also depend on the quality of flat-field correction and the level of signal homogeneity.

Substantial improvement may
be achieved for instruments with non-slit geometry, e.g., integral field units, or
narrowband filters, with large FoV.
For example, for $\Omega=\Omega_{\rm B68}$
the FoV solid angle is larger by a factor of $\approx 1,900$ (compared to the 11'' slit), 
equivalent to an improvement of a factor $\sqrt{1900} \approx 44$ in the SNR.
An alternative avenue is to use space-based observatories, such as the upcoming 
James Webb Space Telescope.
From space, the noise in the IR is much lower, 
and at the same time the O(2) line, which is a factor of 4 brighter than S(0), is accessible (see Table \ref{table}).

\section*{Discussion}
\label{sec: conclusions}

We presented an analytic study of H$_2$ rovibrational line formation produced by penetrating CRs, as well as by the competing processes: H$_2$ formation pumping, and 
UV pumping,
and investigated the conditions required for (a) CRs to dominate line formation, and (b) for the lines to be sufficiently bright to be detected.
We showed that in cold dense clouds, exposed to the mean UV field, 
the 
(1-0)O(2), Q(2), S(0), O(4) line emission is dominated by
CR pumping, and thus detection of these lines may be used to constrain the CRIR.

Whether the lines are excited by CRs, UV, or formation pumping may be determined by the line ratios.
For example, the ratio of 1-0 S(1) to 1-0 S(0) lines is $\approx 2$ for UV excitation \cite[]{Sternberg1988}, and is in the range $3.5-5.6$ for formation pumping  \cite[]{LeBourlot1995}.
On the contrary, for cold clouds excited by CRs, this ratio is predicted to be $\ll 1$ (see Table \ref{table}).

Observations of the H$_2$ lines may be an efficient method to determine the CRIR in dense clouds.
A survey of several clouds in various regions in the Galaxy may reveal the degree of fluctuations in the CRIR, while comparison with the CRIR in diffuse clouds (as probed by chemical tracers, e.g.,  H$_3^+$, ArH$^+$, etc.),  may constrain the  attenuation 
of CRs with cloud depth, and therefore the spectrum of low energy CRs \cite[][]{Padovani2009}.
Such tests may shed light on the nature and formation process of CRs in the Galaxy.

\section*{Methods}
\subsection*{Relative line Brightness}
\label{sub: fi}
The line brightness following CR excitation depend on the excitation probabilities, $p_{u, {\rm (cr)}}$.
We derive $p_{u, {\rm (cr)}}$ and $\varphi$ based on data from
\cite{Gredel1995}, assuming $T=30$ K and electron energy 30 eV.
These authors presented data for the excitation to level $u$ per CR ionization (rather than per CR excitation), denoted $b_u$ - see their Table 2.
Comparing our and their definitions, we get
$p_{u, {\rm (cr)}} =b_u \zeta/\zeta_{ex} = b_u/\varphi$, $\varphi \approx 5.8$.

For H$_2$ formation pumping, we obtain $f_{ul,({\rm f})}$ for each of the three models $\phi=1, 2, 3$ model based on \cite{LeBourlot1995}.
We divided their reported line brightness by $I_{\rm tot, {\rm f}}$, as given by our Eq.~(\ref{eq: I tot form 1}-\ref{eq: I tot form 2}) with $N_{22}=1$, $\chi=0.58$, $\zeta_{-16}=0.1$ appropriate to the assumed values in \cite{LeBourlot1995}, and assuming $\varphi_E = (1.15, 3.5, 1.5)$ for $\phi=(1, 2, 3)$, respectively \cite[]{Black1987}.

\subsection*{Gas temperature}
\label{sub: gas temp}

\ctable[
caption = \mbox{Cosmic-ray pumping - temperature dependence},
label = table2,
pos = t,
]{l l l l l l l}{
\tnote[a]{Based on \cite{Gredel1995} and \cite[][$x_{\rm e}=10^{-6}$]{Tine1997}
}
\tnote[b]{The O(2), Q(2), S(0) and O(4) intensities are weakly sensitive to temperature at low $T$: 
they remain within 3\% for $T<34$ K, and within (10, 20, 30)\% for $T<($41, 50, 58) K, respectively.
}
}{ 
\FL
Transition & $J_u$ & $J_l$ & $\lambda$ & 
\multicolumn{3}{c}{$f_{ul,{\rm (cr)}} \ (\%)$\tmark[a]}
\\
 &  &  & ($\mu$m) & $30$ K\tmark[b] & $100$ K & $300$ K
 \ML
(1-0)O(2) &0 & 2 &2.63 &45.29 & 16.41&9.69  \NN 
(1-0)Q(2) &2 & 2 &2.41 &17.67 & 6.40& 4.31\NN 
(1-0)S(0) &2 & 0 &2.22 &15.99 & 5.79&3.90\NN 
(1-0)O(4) &2 & 4 &3.00 &13.57 & 4.92&3.31\NN
(1-0)Q(1) &1 & 1 &2.41 &0.76 &15.34 &17.76\NN 
(1-0)S(1) &3 & 1 &2.12 &0.71 & 14.37&16.64\NN 
(1-0)O(3) &1 & 3 &2.80&0.64 & 12.97&15.02\NN 
(1-0)Q(3) &3 & 3 &2.42 &0.50 & 10.09&11.67\NN 
(1-0)O(5) &3 & 5 &3.23 &0.28 & 5.66&6.55\LL
}

In the results section, we focused on the low $T \lesssim 50$ K regime, typical of cold molecular cloud interiors.
Here we discuss the case of warmer gas.
We have carried out calculations for the line intensities as a function of temperature based on data from \cite{Gredel1995} and \cite{Tine1997}.
Results for $T=30$, 100 and 300 K are presented in Table \ref{table2}.
As long as $T<60$ K, the spectrum is heavily dominated by the para-H$_2$ lines. In this limit the para-H$_2$ lines remain insensitive to $T$.
This is because at these temperatures the H$_2$ molecules always reside mostly in the ground $(v,J)=(0,0)$ state.
For $T \gtrsim 60$ K, the $(v,J)=(0,1)$ level is sufficiently populated such that CR pumping from this level effectively
excites the $(v,J)=(1,1)$ and $(1,3)$ states, resulting in emission of ortho-H$_2$ lines: S(1), Q(1), Q(3), O(3), and O(5).
While the power in each individual transition is reduced, the total power summed over the lines is conserved.

\subsection*{Variation of $\zeta$ with cloud depth}
\label{sub: zeta var cloud}
In our Eqs.~(\ref{eq: I_tot cr}-\ref{eq: I_ul}) and (\ref{eq: I tot form 1}) we assumed a constant CRIR.
In practice, CRs interact with the gas leading to an attenuation of the CRIR with an increasing gas column.
 For columns $N=10^{20}-10^{25}$ cm$^{-2}$, 
 the $\zeta-N$ relation may be described by a power law,
 \begin{equation}
 \label{eq: varying zeta}
 \zeta(N_{\rm H_2}) = \zeta_0 \left(\frac{N_{\rm H_2}}{N_0} \right)^{-a}
 \end{equation}
with $N_0 = 10^{20}$ cm$^{-2}$, and where the power-index $a$ and the normalization $\zeta_0$ depend on the spectrum of the CRs 
 \cite[][]{Padovani2009}.
To account for a varying $\zeta$, 
 our expressions for the total line brightness should be modified as follows:
 \begin{align}
 \label{eq: varying zeta}
 \zeta N_{\rm H_2} &\rightarrow \int_0^{N_{\rm H_2}} \zeta(N_{\rm H_2}) \mathrm{d} N_{\rm H_2}  = \frac{\zeta_0 N_0}{1-a} \left( \frac{N_{\rm H_2}}{N_0} \right)^{1-a} \\ \nonumber
 &=  \zeta(N_{\rm H_2}) N_{\rm H_2}  \  \frac{1}{1-a}
 \end{align}
where we solved the integral assuming Eq.~(\ref{eq: varying zeta}) with $a \neq  1$ (for the four CR spectra considered by \cite{Padovani2009}, $a=0.021, 0.423, 0.04, 0.805$).

Eq.~(\ref{eq: varying zeta}) shows that even in the case of a varying CRIR, our Eqs.~(\ref{eq: I_tot cr}-\ref{eq: I_ul}, \ref{eq: I tot form 1})  still provide an excellent approximation for the line brightness, but with $\zeta$ representing the CRIR in cloud interior.
The factor $1/(1-a)$ approaches unity 
for relatively flat spectra (i.e., models 1 and 2 in \cite{Padovani2009}), and the brightness is then independent of the spectrum shape.
If H$_2$ line observations are further combined  with 
additional observations of the CRIR in diffuse cloud regions (e.g., with ArH$^+$, OH$^+$, H$_2$O$^+$\cite[][]{Neufeld2017b}),
the CR attenuation may be obtained, constraining the CR spectrum.

\subsection*{Line brightness dependence on $\zeta$}
Our Eq.~(\ref{eq: I_tot cr}) suggests that the line emission is linear in $\zeta$.
This relation holds as long as $\zeta$ is not too high.
With increasing $\zeta$ both gas temperature increases (which affect the excitation pattern), and more importantly, the electron fraction, $x_{\rm e}$ increases.
When $x_{\rm e} \gtrsim 10^{-4}$, coulomb energy loses become substantial and line excitation is quenched (see Tables 2-3 in \cite{Tine1997}).
 However, this requires extreme CRIR, such that the gas is no longer molecular \cite[]{Bayet2011a, Bialy2015a, LePetit2016}.

\subsection*{Collisional de-excitation}
 \label{sub col de-ex}
At sufficiently high density, collisional deexcitation (by thermal H$_2$, H, etc.) dominates over radiative decay, and the line emission is quenched.
The critical density at which collisional de-excitation equals radiative decay is
 $n_{{\rm crit}}(T)=A_{ul}/(x_{\rm col}k_{ul\downarrow}(T))$, where  $A_{ul}$ is the Einstein coefficient for spontaneous emission, $k_{ul\downarrow}(T)$ is the 
 collisional rate coefficient,  and
 $x_{\rm col}=n_{\rm col}/n$   is the fractional abundance of the collision partner.
 
Let us estimate the critical density for $v=1-0$ deexcitation in cold-dense clouds.
For the $v=1 \rightarrow 0$,   $A_{ul}\approx (2-8) \times 10^{-7}$ s$^{-1}$ (see Table \ref{table}).
The rate coefficients at $T = 100$ K are of order of
 $k_{\downarrow} \approx  10^{-13}$ cm$^3$ s$^{-1}$ and $k_{\downarrow} \approx 5 \times 10^{-18}$ cm$^3$ s$^{-1}$, for collisions with H and H$_2$, respectively \cite[][]{Lique2015, Flower1999}.
 In cloud interiors, the H/H$_2$ ratio is set by the balance of H$_2$ ionization by CRs and H$_2$  formation via dust catalysis. This gives $x_{\rm H}/x_{\rm H_2} \approx \zeta/(R n) \approx 3.3 \times 10^{-5} \zeta_{-16}/(n_5 T_2^{0.5})$ where $R \approx 3 \times 10^{-17} T_2^{0.5}$ cm$^3$ s$^{-1}$ is the H$_2$ formation rate coefficient, $T_2 \equiv T/(100 \ {\rm K})$ and $n_5 \equiv n/(10^{5} \ {\rm cm^{-3}})$ (for more details see \S 4.1 in \cite{Bialy2015a}).
Hydrogen nucleus conservation ($x_{\rm H}+2 x_{\rm H_2} \simeq 1$) then implies $x_{\rm H_2} \approx 0.5$, and $x_{\rm H}  \approx 1.7 \times 10^{-5} \zeta_{-16}/n_5$.
For $T=100$ K we obtain 
 $n_{{\rm crit}}$ of order of $10^{11}$ cm$^{-3}$, both for collisions with H and H$_2$.
 In practice, the temperature in cloud cores is typically lower than 100 K leading to even lower collisional rates ($k_{\downarrow}$), and thus even higher critical densities.
In conclusion,  for the typical temperatures and densities in cold clouds ($T < 100$ K, $n \approx 10^4-10^6$ cm$^{-3}$), radiative decay strongly dominates over collisional deexcitation from the $v=1$ levels.

\subsection*{Thermal excitation}
 \label{sub thermal ex}
 In our model we ignored thermal (collisional) excitation.
 Although excitations by the non-thermal CRs occurs rarely  (at a rate $\sim \zeta$), thermal excitation at $T\lesssim 100$ K is extremely negligible.
 The thermal excitation rate is $q = x_{\rm col} nk_{ul\downarrow} \exp(-\Delta E_{ul}/(k_B T)) g_{u}/g_{l}$, 
where $g_{u},g_{l}$ are the quantum weights of the levels.
While $x_{\rm col} k_{ul\downarrow} \sim 10^{-18} - 10^{-16}$ cm$^{3}$ s$^{-1}$, 
the exponential factor is $\sim 10^{-22}$ ($\Delta E_{ul}/k_B \approx 5500$ K). Thus, thermal excitation is negligible. 

\subsection*{Exposure Time Calculator}
For our estimation of the signal to noise ratio per pixel per hour integration, we used the X-shooter exposure time calculator provided in
\href{https://www.eso.org/observing/etc/bin/gen/form?INS.NAME=X-SHOOTER+INS.MODE=spectro
}{https://www.eso.org/observing/etc/bin/gen/form?INS.NAME=X-SHOOTER+INS.MODE=spectro},
with the following setup.
Emission Line:
Lambda (2223.2, 2413.3) nm for S(0), Q(2) respectively.
Flux $(0.0089, 0.0098) \times 10^{-16}$ erg s$^{-1}$ cm$^{-2}$ arcsec$^2$ for S(0), Q(2) respectively. FWHM=0.093 nm (appropriate for a single spectral resolution element of the 0''4 slit).
Spatial distribution, Extended source.
Moon FLI 0.5, Airmass 1.5, PWV 30mm, Turbulence Category 70\%.
NIR slit width 0''.4, 
DIT=900 s, NDIT=4.

\vspace*{0.3cm}
{\bf Acknowledgements:}
SB thanks Alyssa Goodman, David Neufeld, Amiel Sternberg, Oren Slone, Brian McLeod and Igor Chilingaryan for fruitful discussions. 

\vspace*{0.3cm}
{\bf Author contribution:}
The author carried out all the analytic derivations, numerical computations,  and manuscript writing.

\vspace*{0.3cm}
{\bf Competing Interests:}
The author declares no competing interests.

\vspace*{0.3cm}
{\bf Data Availability:}
The author declares that all data supporting the findings of this study is available within the paper.

\end{document}